\documentclass[preprint,aps,amsmath,amssymb,superscriptaddress,floatfix]{revtex4-2}

\usepackage[utf8]{inputenc}
\usepackage{setspace}
\usepackage{graphicx}
\usepackage{color}
\usepackage{soul}
\usepackage{hyperref}
\usepackage{miller}
\usepackage{pdfpages}
\usepackage{bbold}
\usepackage{gensymb}
\usepackage{orcidlink}

\newcommand{\simlt}
      {\ifmmode       { \raisebox{-.8em}{$<$}\atop\sim}
         \else        {$\raisebox{-.8em}{$<$}\atop\sim$}
      \fi}

\makeatletter
\def\@hangfrom@section#1#2#3{\@hangfrom{#1#2}#3}
\def\@hangfroms@section#1#2{#1#2}
\makeatother

\makeatletter
\AtBeginDocument{\let\LS@rot\@undefined}
\makeatother

\usepackage[version=4]{mhchem}

\usepackage[dvipsnames]{xcolor}

\begin{document}


\title{Revealing Hund superdispersion with tunneling spectroscopy}

\author{Luke C. Rhodes\,\orcidlink{0000-0003-2468-4059}}
\email[]{lcr23@st-andrews.ac.uk}
\affiliation{SUPA, School of Physics and Astronomy, University of St Andrews, North Haugh, St Andrews, KY16 9SS, United Kingdom}

\author{Fabian B.~Kugler\,\orcidlink{0000-0002-3108-6607}}
\email[]{fkugler@thp.uni-koeln.de}
\affiliation{Institute for Theoretical Physics, University of Cologne, 50937 Cologne, Germany}

\author{Olivier~Gingras\,\orcidlink{0000-0003-3970-6273}}
\email[]{ogingras@flatironinstitute.org}
\affiliation{Center for Computational Quantum Physics, Flatiron Institute, 162 5th Avenue, New York, New York 10010, USA}
\affiliation{Université Paris-Saclay, CNRS, CEA, Institut de physique théorique, 91191, Gif-sur-Yvette, France}

\author{Carolina Marques\,\orcidlink{0000-0002-3804-096X}}
\affiliation{SUPA, School of Physics and Astronomy, University of St Andrews, North Haugh, St Andrews, KY16 9SS, United Kingdom}

\author{Edgar Abarca Morales\,\orcidlink{0000-0002-7714-8228}}
\affiliation{Max Planck Institute for Chemical Physics of Solids, Nöthnitzer Strasse 40, 01187 Dresden, Germany}

\author{Phil D.C. King\,\orcidlink{0000-0002-6523-9034}}
\affiliation{SUPA, School of Physics and Astronomy, University of St Andrews, North Haugh, St Andrews, KY16 9SS, United Kingdom}

\author{Antoine Georges\,\orcidlink{0000-0001-9479-9682}}
\affiliation{Collège de France, Université PSL, 11 Place Marcelin Berthelot, 75005, Paris, France}
\affiliation{Center for Computational Quantum Physics, Flatiron Institute, 162 5th Avenue, New York, New York 10010, USA}
\affiliation{Centre de Physique Théorique, Ecole Polytechnique, CNRS, Institut Polytechnique de Paris, 91128, Palaiseau Cedex, France}
\affiliation{DQMP, Université de Genève, 24 Quai Ernest Ansermet, 1211, Genève, Suisse}

\author{Peter Wahl\,\orcidlink{0000-0002-8635-1519}}
\email[]{gpw2@st-andrews.ac.uk}
\affiliation{SUPA, School of Physics and Astronomy, University of St Andrews, North Haugh, St Andrews, KY16 9SS, United Kingdom}
\affiliation{Physikalisches Institut, Universität Bonn, Nussallee 12, 53115 Bonn, Germany}

\date{\today}

\maketitle

\textbf{In cuprate superconductors, electron-electron repulsion results in characteristic spectroscopic features known as `waterfalls', where the sharp quasiparticle dispersion transitions into broad Hubbard bands. However, in multi-orbital systems, the additional Hund coupling results in behavior that defies the conventional Mott--Hubbard paradigm, creating qualitatively distinct `superdispersive' features in the spectral function. Here, we use tunneling spectroscopy to reveal this signature of Hund physics in \ce{Sr2RuO4}. By combining density functional theory, dynamical mean-field theory, and continuum local density of states calculations, we show that the experimental features are in excellent agreement with theoretical predictions and intimately linked to the non-monotonous energy dependence of the real part of the self-energy in a Hund metal. Our results provide direct experimental evidence for Hund-induced spectroscopic features and open a new route to probing correlation effects in quantum materials.}


\vspace{5mm}
\large\noindent\textbf{Introduction}
\vspace{2mm}
\normalsize

The physics of Hund coupling continues to challenge our understanding of correlated electron systems. Far from being a simple extension of Mott physics, it plays a central role in multi-orbital quantum materials, organizing a rich landscape of emergent phenomena. These include spin-orbital separation, enhanced electronic correlations coexisting with bad metallicity, and orbital-selective electronic orders~\cite{yin_kinetic_2011, georges_strong_2013, deMedici2017_HundsMetals, Kugler2019, Fanfarillo_2020_Synergy, PhysRevResearch.6.023124, georges_hund-metal_2024}. Effects of Hund coupling occur simultaneously in high-energy spectral features associated with atomic multiplet configurations and in the low-energy quasiparticle dynamics showing dramatic band renormalizations which have a highly non-trivial dependence on interaction strength and orbital filling~\cite{de_medici_janus-faced_2011, de_medici_hunds_2011}.

Hund physics manifests itself most prominently when the ground-state configuration of the active atomic shell has both spin and orbital degeneracy in the atomic limit (i.e.\ in the absence of electronic hopping). Hund coupling then severely modifies the phase-space for hopping inducing strong electronic correlations in itinerant metals which are not close to a Mott insulator~\cite{georges_hund-metal_2024}. Thus, in contrast to Mott insulators where correlations stem from charge blocking, correlations in Hund metals originate in spin and orbital blocking~\cite{yin_kinetic_2011,georges_strong_2013,fanfarillo_electronic_2015, STADLER2019365}.

Unraveling the experimental signatures and physical effects associated with Hund coupling is essential to assess the reliability of first-principles approaches to strong electronic correlations and to predict and engineer quantum materials in which Hund physics drives emergent behavior.

A distinctive feature of the correlations in the single-orbital Hubbard model are spectroscopic `waterfall' features, i.e.\ an extended energy range connecting the low-energy quasiparticle band and the high-energy Hubbard bands. The resulting single-particle spectral function exhibits vertical features in the energy-momentum plane often stretching over several eV~\cite{Si_Closing_2024, krsnik_local_2025}. These have been seen extensively in the cuprate superconductors in photoemission~\cite{Moritz_2009}. 
 
\begin{figure}
    \centering
    \includegraphics[width=\linewidth]{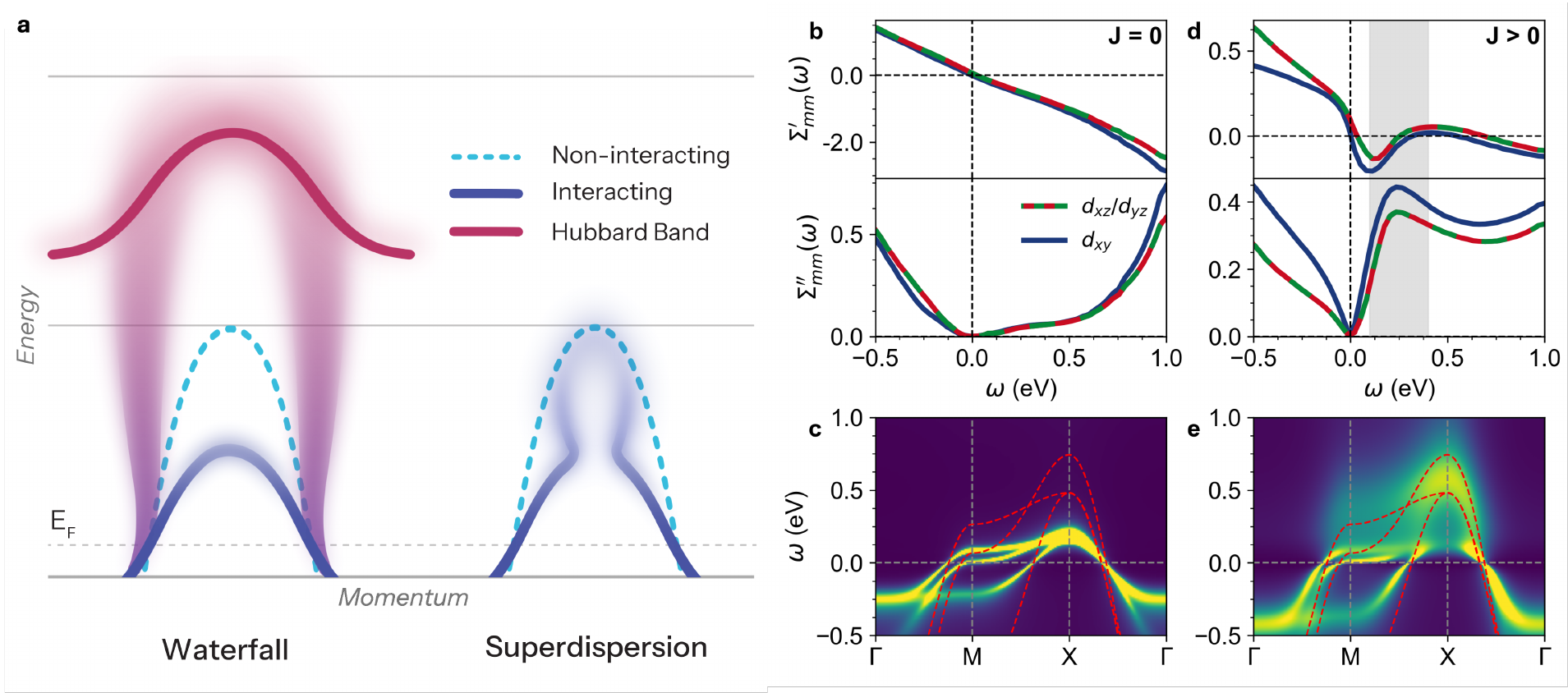}
    \caption{
        \textbf{Waterfalls and superdispersive spectral features.}
        (a)~Schematic of a waterfall (left), where vertical waterfall-like features connect the 
        quasiparticle band with the Hubbard band, and the superdispersion found in a Hund metal (right), where a reversal of the dispersion originates from  successive renormalization and `unrenormalization' of the non-interacting bands.
        (b,d)~Diagonal elements of the real ($\Sigma'_{mm}(\omega)$) and imaginary ($\Sigma''_{mm}(\omega)$) parts of the self-energy for \ce{Sr2RuO4}.
        In (b), we take $(U,J)=(4.5,0)$~eV, chosen to produce similar quasiparticle weights as for the physical case $(U,J)=(2.3,0.4)$~eV in (d).
        (c,e)~Spectral functions of bulk \ce{Sr2RuO4} along the $\Gamma$-$M$-$X$-$\Gamma$ path incorporating the self-energies of (c,d), respectively. The red dashed lines indicate the non-interacting band structure. See Suppl. Sect. S1 and Fig. S1 for the definition of the high-symmetry points.  
        } 
    \label{fig:Waterfalls}
\end{figure}
 
In contrast, the spin and orbital blocking inherent to Hund metals can generate regimes in which the interaction-induced renormalization of the band structure has a non-monotonous dependence on energy. The resulting interacting bands have energy ranges that are both `renormalized', where the group velocity is reduced compared to the non-interacting band structure, as well as energy ranges that are `unrenormalized', where the group velocity is larger than the non-interacting band structure. This leads to low-energy features in the spectral functions where the dispersion is vertical and, in a limited energy range, even reverses direction, an effect we call `superdispersion'. Figure~\ref{fig:Waterfalls}(a) illustrates the two cases of a waterfall, as found in the single-orbital Hubbard model, and the superdispersion predicted for Hund metals which we discuss here. 

The occurrence of a superdispersion has been theoretically predicted~\cite{Stricker2014} from dynamical mean-field theory (DMFT) calculations in the iconic Hund metal \ce{Sr2RuO4}~\cite{Mravlje_Coherence_2011}. Whereas the electronic structure of \ce{Sr2RuO4} is exceptionally well characterized by quantum oscillations~\cite{mackenzie_quantum_1996} and angle-resolved photoemission spectroscopy (ARPES)~\cite{Mravlje_Coherence_2011, tamai_high-resolution_2019}, finding signatures of superdispersion has been experimentally challenging as they are predicted to occur in the unoccupied electronic states, which many conventional probes cannot directly access. 

Here, we report observation of an experimental signature of such a Hund superdispersion in the unnocupied density of states via tunneling spectroscopy measurements. By combining \textit{ab initio} modeling of the surface electronic structure of \ce{Sr2RuO4} with DMFT and continuum local density of states (cLDOS) calculations, we show that the signature of Hund superdispersion predicted by DMFT is directly seen in the experimental measurements. This finding and methodological development forms the basis of a new framework to explore the physics of correlated materials and enable quantitative tests for predictions of DMFT calculations.

\vspace{5mm}
\large\noindent\textbf{Superdispersive single-particle excitations}
\vspace{2mm}
\normalsize

To understand the Hund superdispersion, we begin by introducing the single-particle spectral function:
\begin{equation}
    A(\mathbf{k},\omega) = -\frac{1}{\pi} \mathrm{Im}\,\text{Tr}
    \left[
    \omega\cdot\mathbb{1}-\hat{H}_0(\mathbf{k})-\hat{\Sigma}(\textbf{k},\omega) 
    \right]^{-1}.
    \label{Eq:Akw}
\end{equation}
Here, $\hat{H}_0$ is the non-interacting Hamiltonian and $\hat{\Sigma}$ the self-energy encoding the electronic correlations. The `hat' symbol indicates that both quantities are matrices with a size given by the number of bands or Wannier orbitals retained in the description of the electronic structure.  In this work, $\hat{H}_0(\textbf{k})$ is based on the bulk Ru$~4d-t_{2g}$ orbitals obtained from density functional theory (DFT) calculations (see Methods for details) of \ce{Sr2RuO4}.

Being interaction driven, signatures of Hund physics can be understood by inspecting the self-energy. The imaginary part of the self-energy $\hat{\Sigma}''$ encodes the finite lifetime of single-particle excitations and thus blurs the spectrum away from the Fermi energy. The real part $\hat{\Sigma}'$ determines the dispersion $\omega(\mathbf{k})$ of excitations, which follow from the pole equation:
\begin{equation}
\label{Eq:QPdisp}
\mathrm{det} \left[
\omega(\mathbf{k})\cdot\mathbb{1}-\hat{H}_0(\mathbf{k})-\hat{\Sigma}'(\mathbf{k},\omega(\mathbf{k})) 
\right] = 0.
\end{equation}
The solutions provide the renormalized band energies $\omega(\textbf{k}) \to E_\nu(\mathbf{k})$.
In fact, if we work in the band basis, where $\hat{H}_0(\mathbf{k}) = \mathrm{diag}(E^0_\nu(\mathbf{k}))$, and neglect the interband elements of the self-energy, ${\Sigma}'_{\nu\nu'}(\mathbf{k},\omega) \approx {\Sigma}'_{\nu\nu}(\mathbf{k},\omega) \delta_{\nu\nu'}$, we readily find
\begin{equation}
    \label{Eq:QPdisp2}
    E_\nu(\mathbf{k}) \approx E^0_\nu(\mathbf{k}) + {\Sigma}_{\nu\nu}'(\mathbf{k},E_\nu(\mathbf{k})).
\end{equation}
The velocity of single-particle excitations, $\mathbf{v}_\nu(\mathbf{k}) = \nabla_{\mathbf{k}} E_\nu(\mathbf{k})$, is then found from Eq.~\eqref{Eq:QPdisp2} as
\begin{equation}
    \label{Eq:QPvel}
    \mathbf{v}_\nu(\mathbf{k}) \approx \frac{\mathbf{v}^0_{\nu}(\mathbf{k}) + \nabla_{\mathbf{k}} \Sigma_{\nu\nu}'(\mathbf{k},\omega)}{1-\partial_\omega \Sigma_{\nu\nu}'(\mathbf{k},\omega) }\bigg|_{\omega=E_\nu(\mathbf{k})},
\end{equation}
where $\mathbf{v}^0_\nu(\mathbf{k}) = \nabla_{\mathbf{k}} E^0_\nu(\mathbf{k})$.
Through the denominator in Eq.~\eqref{Eq:QPvel}, the frequency dependence of $\hat{\Sigma}'$ induces a potentially drastic renormalization of the band velocities.

As detailed in the Methods section, we compute the self-energy of \ce{Sr2RuO4} using DMFT. Within DMFT, $\hat{\Sigma}$ is computed non-perturbatively, assuming it is local (i.e.\ momentum independent) when expressed in the basis of local (Wannier) orbitals: ${\Sigma}_{mm'} (\textbf{k},\omega) \approx {\Sigma}_{mm'}(\omega)$~\cite{georges_dynamical_1996}. It acquires a significant momentum dependence in the band basis due to spin-orbit coupling, as discussed in Ref.~\citenum{tamai_high-resolution_2019}. 

To explore  different correlation effects on the spectral function, we initially use a self-energy calculated at $J=0$, thus ignoring Hund coupling and considering only the screened Coulomb repulsion $U$. As a consequence, ${\Sigma}_{mm}'(\omega)$ shows a monotonous, almost linear decrease in $\omega$ within the relevant $\pm1$~eV energy window (Fig.~\ref{fig:Waterfalls}(b)). The corresponding spectral function in Fig.~\ref{fig:Waterfalls}(c) displays excitations with smaller velocities than in DFT. This reduction of the band velocities is the well-known renormalization observed in correlated materials.

In contrast, if we include finite $J$ in the DMFT calculation, we find the self-energy presented in Fig.~\ref{fig:Waterfalls}(d) for which ${\Sigma}'_{mm}(\omega)$ becomes non-monotonous and exhibits a pronounced dip at energies slightly above the Fermi level~\cite{Mravlje_Coherence_2011,Stricker2014}. The local minimum is at +100~meV for the $d_{xz/yz}$ orbitals, and +125~meV for the $d_{xy}$ orbital, and is followed by a positive slope of ${\Sigma}'_{mm}(\omega)$ between 100 and 300~meV above the Fermi level, as indicated by the grey shaded region in Fig.~\ref{fig:Waterfalls}(d). 
According to Eq.~\eqref{Eq:QPvel}, $\partial_\omega{\Sigma}'_{mm}(\omega)>0$ leads to band velocities larger than in DFT---an effect coined `unrenormalization'~\cite{Mravlje_Coherence_2011, PhysRevResearch.6.023124}. 
Intriguingly, $\partial_\omega\hat{\Sigma}'_{mm}(\omega)$ can exceed unity, yielding single-particle excitations that disperse infinitely fast and even reverse their dispersion locally---a counter-intuitive effect responsible for superdispersive physics.
In the single-orbital Hubbard model, similar $\partial_\omega{\Sigma}'_{mm}(\omega)>0$ effects are seen; however, they occur at higher ($\sim \mathrm{eV}$) energies at the formation of the Hubbard bands and thus lose the connection to the uncorrelated band structure~\cite{krsnik_local_2025}. This is not the case for Hund metals, where the spectral function is still directly tied to the single-particle band structure, as shown in Fig.~\ref{fig:Waterfalls}(e). Here, the underlying single-particle excitations stay in one-to-one correspondence to the non-interacting band structure, making superdispersive single-particle excitations a distinct feature of Hund metals.

Previous theory works have shown that the local unrenormalization which leads to the superdispersion in \ce{Sr2RuO4} is induced by Hund coupling $J$ and that the dip in ${\Sigma}'_{mm}(\omega)$ disappears when $J$ is suppressed~\cite{kugler_strongly_2020,PhysRevResearch.6.023124,PhysRevLett.125.166401}. The non-monotonous behavior of $\Sigma'_{mm}(\omega)$ can be understood as an energy-dependent accumulation or depletion of spectral weight at certain energies~\cite{Stadler2015,Kugler2019,kugler_strongly_2020,Kugler2024}, leaving a trace in the local density of states.
It thus forms a unique spectroscopic signature of Hund physics that can be probed experimentally. 

\vspace{5mm}
\large\noindent\textbf{The self-energy at the surface of \ce{Sr2RuO4}}
\vspace{2mm}
\normalsize

\begin{figure}
    \centering
    \includegraphics[width=0.99\linewidth]{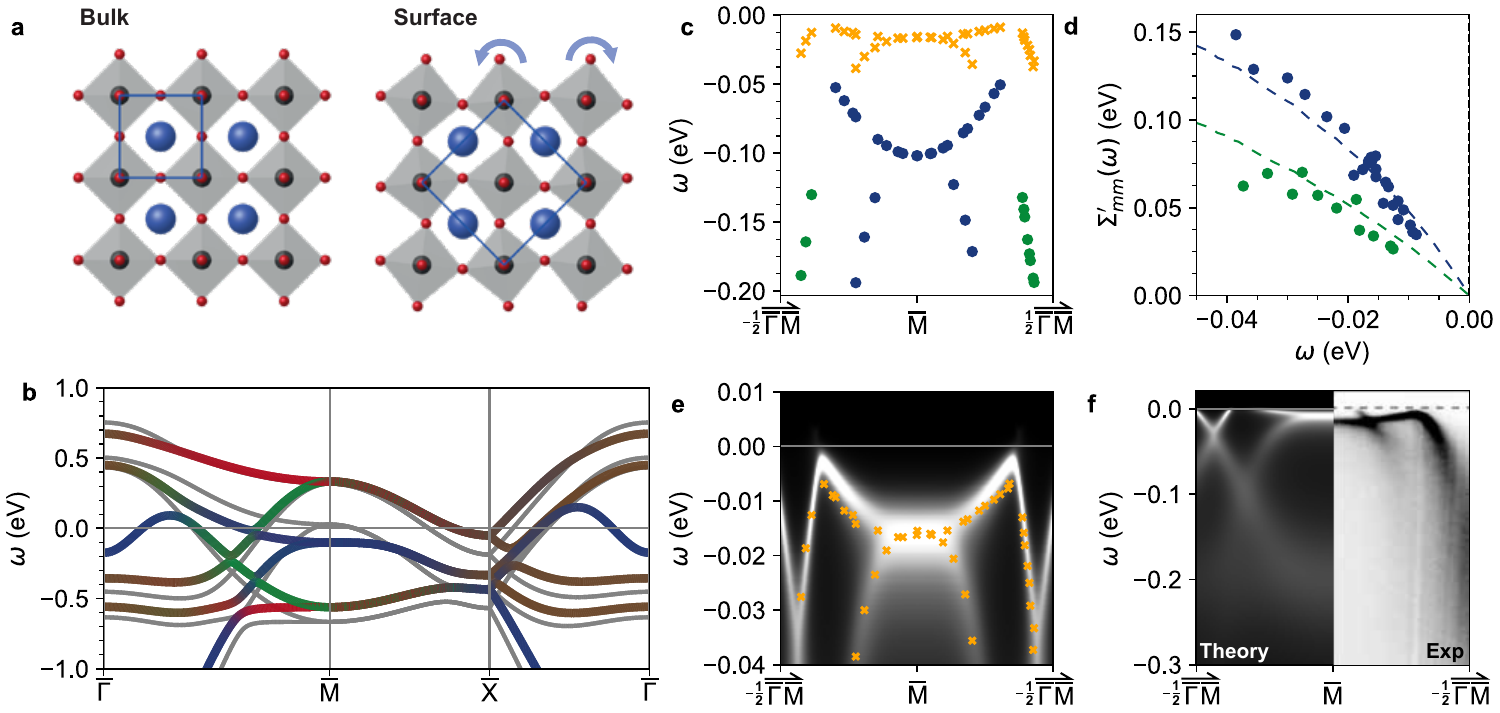}
    \caption{
        \textbf{Extracting the self-energy at the surface of \ce{Sr2RuO4}.}
        (a) $ab$-plane projection of \ce{Sr2RuO4} for the bulk unrotated structure (left) and surface rotated structure (right). The surface has a 9$^{\circ}$ rotation of the \ce{RuO6} octahedral (grey squares) compared to the bulk. 
        (b)~Band structure of the $9^{\circ}$ rotated surface of \ce{Sr2RuO4} from DFT, with dominant orbital characters shown by red, green and blue for the $d_{xz}$, $d_{yz}$ and $d_{xy}$ orbitals, respectively. The thin grey bands indicate the electronic structure of the unrotated two-atom unit cell of \ce{Sr2RuO4} (equivalent to the bulk structure) for comparison. 
        (c)~Comparison of ARPES data points (orange crosses) around the $M$ point, extracted from Ref.~\citenum{Morales_Hierarchy_2023,chandrasekaran_engineering_2024}, with the DFT eigenvalues and dominant orbital characters plotted at the same momentum. 
        (d)~Real part of the diagonal part of the self-energy extracted from ARPES measurements by taking the difference in energy between the ARPES data and the non-interacting band dispersion. The dashed lines are the orbital-resolved diagonal elements of the real parts of the DMFT bulk self-energy. A rigid offset to the extracted points and calculated self-energy was applied to enforce ${\Sigma'(\omega=0)}=0$ for comparison with Ref.~\citenum{tamai_high-resolution_2019}.
        (e, f)~Comparison between theory (surface band structure with DMFT bulk self-energy) and ARPES around the $M$ point, in (e) a low-energy window (dispersion extracted from ARPES data in orange crosses) and (f) a large-energy window (ARPES data from Ref.~\citenum{chandrasekaran_engineering_2024}).
    } 
    \label{fig:ARPES}
\end{figure}

Low-temperature scanning tunneling microscopy and spectroscopy (STM/STS) provide direct access to both occupied and unoccupied states, with atomic spatial resolution and energy resolution limited practically only by temperature. In principle, this enables us to probe the electronic structure of the unoccupied states. However, interpreting STM measurements in terms of the underlying electronic structure is considerably more involved than for the conventional method of ARPES, often necessitating sophisticated numerical modeling~\cite{wahl_calcqpi_2025}. This challenge is further compounded by the well-documented surface reconstruction of \ce{Sr2RuO4}~\cite{matzdorf_ferromagnetism_2000,matzdorf_surface_2002, chandrasekaran_engineering_2024}, rigidly rotating the \ce{RuO6} octahedra as sketched in Fig.~\ref{fig:ARPES}(a) (which also suppresses superconductivity at the surface~\cite{marques_magneticfield_2021,profe_magic_2024,Valadkhani_why_2024}). Thus, before we can explore the unoccupied correlated electronic structure of \ce{Sr2RuO4} for signatures of the Hund superdispersion, we first have to understand what changes are induced by the surface modifications in the spectral function and in the self-energy. 

The self-energy of the bulk bands of \ce{Sr2RuO4} in the occupied region has been established in exquisite details from both DMFT and experiment~\cite{tamai_high-resolution_2019}. In the latter case, $\hat{\Sigma'}$ was extracted via Eq.~\eqref{Eq:QPdisp2}, taking $E_\nu(\mathbf{k})$ from the photoemission peak positions.
More recently, high-resolution ARPES measurements of the reconstructed surface~\cite{Morales_Hierarchy_2023,chandrasekaran_engineering_2024} now enable a similar analysis for the surface electronic structure. The DFT-derived non-interacting bands of the surface electronic structure are shown in Fig.~\ref{fig:ARPES}(b). One of the main differences compared to the bulk electronic structure (grey bands in Fig.~\ref{fig:ARPES}(b)) is that the van Hove singularity (vHs) at the $M$-point shifts from above the Fermi level in the bulk to below the Fermi level at the surface~\cite{matzdorf_surface_2002,marques_magneticfield_2021}. 

To extract $\hat{\Sigma}'$ at the surface, we have performed a similar analysis to what has been done for the bulk. We first show in Fig.~\ref{fig:ARPES}(c) the dispersion close to the vHs extracted from ARPES (orange crosses) with the non-interacting dispersion plotted at the same momenta for the $d_{xy}$ dominated band (blue) and for the $d_{xz/yz}$ dominated band (green). We consider an energy-momentum window with two bands separated by dominant orbital characters and thus obtain two distinct branches of $\hat{\Sigma}'(\omega)$, plotted in Fig.~\ref{fig:ARPES}(d). This analysis shows that: (1)~$\partial_\omega\hat{\Sigma}'<0$ for $\omega<0$, as in the generic case (see Fig.~\ref{fig:Waterfalls}(d)), and (2)~bands with dominant $d_{xz}/d_{yz}$ orbital character have a significantly smaller slope $|\partial_\omega\hat{\Sigma}'|$ compared to bands with dominant $d_{xy}$ orbital character, showing that correlations are larger in the $d_{xy}$ band. This picture is fully consistent with the analysis of the bulk bands~\cite{tamai_high-resolution_2019}. In fact, overlaying $\hat\Sigma'(\omega)$ obtained from DMFT calculations for the bulk of \ce{Sr2RuO4} onto this analysis (dashed lines in \ref{fig:ARPES}(d)) reveals excellent agreement between the two, with no additional fitting parameters. Further details for this extaction methodology can be found in the Suppl. Sect. S2 and Fig. S2.

In Fig.~\ref{fig:ARPES}(e), we plot the spectral function around the $\bar{M}$-point calculated with the surface electronic structure from Fig.~\ref{fig:ARPES}(b) and the bulk DMFT self-energy from Fig.~\ref{fig:Waterfalls}(d) which also reproduces well the band positions from the experimental ARPES data (orange crosses). Additionally, an increased broadening of the spectral function with increasing binding energy, represented by $\hat{\Sigma}''(\omega)$, is in equally good qualitative agreement with that observed in the ARPES dispersion in a larger energy range (Fig.~\ref{fig:ARPES}(f)). Together, this indicates that bulk DMFT calculations of the self energy provide an adequate description of the self-energy also for the surface. This motivates us to embed the bulk DMFT self-energy into our surface \textit{ab initio} electronic structure also for the subsequent analysis.

\vspace{5mm}
\large\noindent\textbf{Realistic modelling of STM with DMFT}

\vspace{2mm}
\normalsize

\begin{figure}
    \centering
    \includegraphics[width=0.99\linewidth]{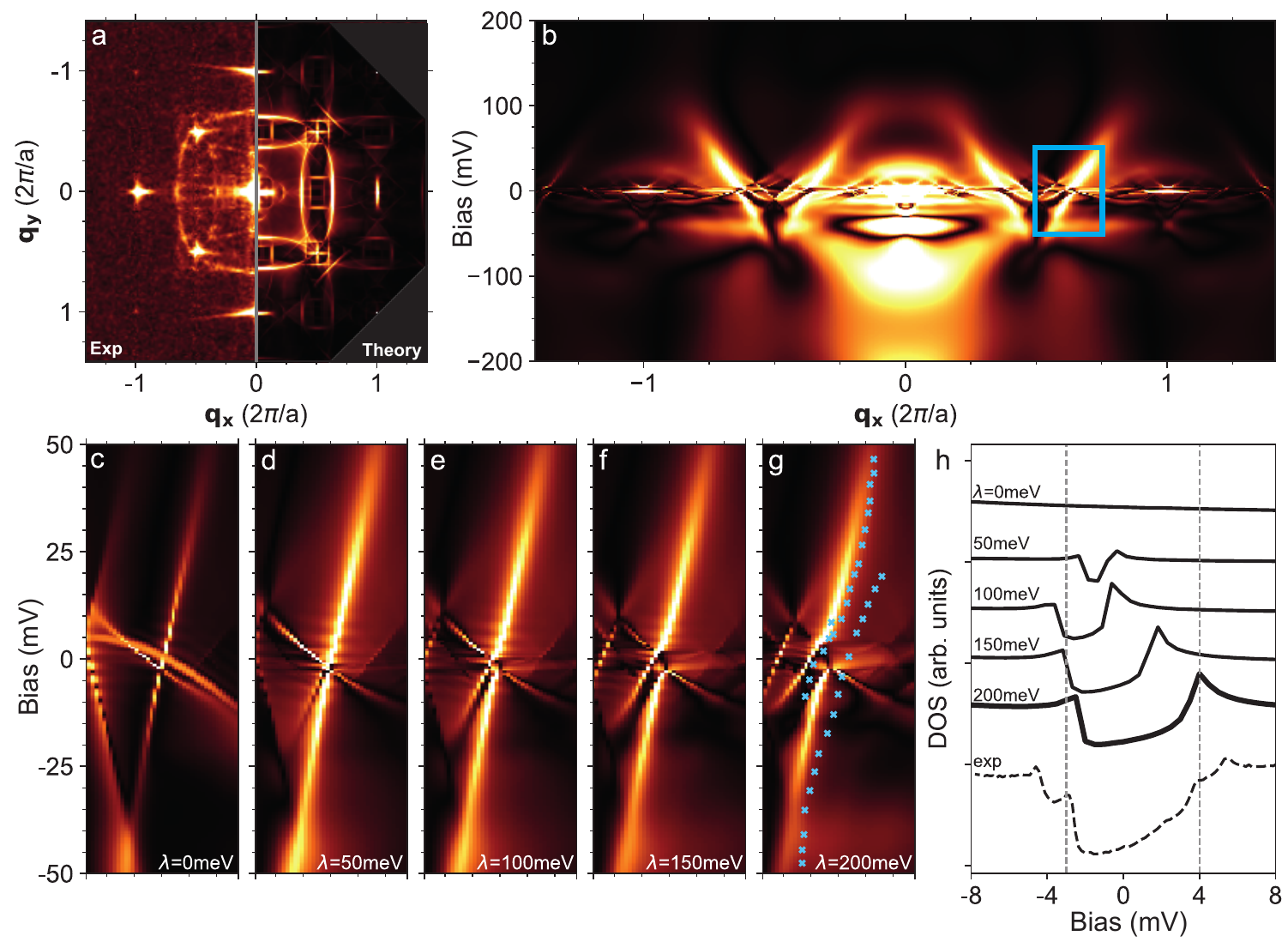}
    \caption{
        \textbf{Correlation-enhanced spin-orbit coupling at the surface of \ce{Sr2RuO4}.}
        (a)~Comparison of the experimental QPI (left-hand side, taken from Ref.~\citenum{kreisel_quasi-particle_2021}) with cQPI+DMFT calculations (right-hand side) at $4.2~\mathrm{mV}$ bias. 
        (b)~Energy dispersion along the $\mathbf{q_y}=0$ line shown in a large energy range, revealing the modification of the scattering intensity due to self-energy effects with significant broadening away from the Fermi energy. 
        (c-g)~Closeup of the blue box in (b) for models calculated with different strengths of the spin-orbit coupling $\lambda$ (specified in the lower right corner). A hybridization gap opens around the Fermi level producing two split branches of the dispersion, as seen in experiment (blue crosses, taken from Ref.~\citenum{wang_quasiparticle_2017}).
        (h)~The low-energy continuum density of states for models with different spin-orbit coupling strengths $\lambda$, along with the experimental spectrum measured in STM ($I_{\mathrm s}=500\mathrm{pA}$, $V_{\mathrm s}=8\mathrm{mV}$, $V_{\mathrm L}=155\mathrm{\mu V}$, $T=56\mathrm{mK}$, see Methods, taken from Ref.~\citenum{marques_magneticfield_2021}). We find that a value of $\lambda=200$~meV (approximately twice the DFT value~\cite{Zhang2016,Kim2018,tamai_high-resolution_2019,Linden2020}) matches the spin-orbital gap seen in the experimental differential-conductance spectrum $g(V)$.
    }
    \label{fig:QPI}
\end{figure}

A detailed comparison of the correlation-induced changes to the electronic structure with STM measurements requires realistic modeling of tunneling spectra. To this end, we introduce the self-energy obtained from DMFT into the continuum LDOS framework~\cite{wahl_calcqpi_2025} and compare the simulated  quasiparticle interference (QPI) with experimental data. Although STM and QPI measurements on \ce{Sr2RuO4} have been performed previously~\cite{wang_quasiparticle_2017,marques_magneticfield_2021,kreisel_quasi-particle_2021,chandrasekaran_engineering_2024}, the comparison with calculations have only considered correlation effects at a phenomenological level through a constant band renormalization of the bare electronic structure~\cite{kreisel_quasi-particle_2021,chandrasekaran_engineering_2024}. Such analysis neglects any energy dependence of the self-energy, which becomes crucial at higher energies and for exploring the Hund superdispersion or, indeed, waterfall phenomena in a Mott--Hubbard system. By performing this full ab-initio + DMFT + continuum LDOS calculation, we find that the calculated low-energy QPI close to the Fermi level agrees very well with the experimental QPI, as shown in Fig.~\ref{fig:QPI}(a). In the energy-dependent QPI, Fig.~\ref{fig:QPI}(b), we observe significant broadening of the QPI signatures and an almost complete suppression of sharp QPI features outside a window of $\pm50\mathrm{meV}$, induced by ${\Sigma}_{mm}''(\omega)$.

Within a $\pm50\mathrm{meV}$ energy window, the improved agreement allows us to identify the role of correlation-enhanced spin-orbit coupling (SOC) at the surface of \ce{Sr2RuO4}. It is known that electronic correlations enhance the SOC value $\lambda$ from DFT by a factor of roughly two in the bulk of \ce{Sr2RuO4}~\cite{tamai_high-resolution_2019}. Adding SOC to the tight-binding model, we can study how the QPI changes for different values of $\lambda$ and thereby extract its impact on the low-energy experimental observables. Focusing on the \textbf{q}-space region indicated by the blue box in Fig.~\ref{fig:QPI}(b), we plot the QPI dispersion for different values of $\lambda$ in Fig.~\ref{fig:QPI}(c-g). Comparing with experimental data (blue squares in Fig.~\ref{fig:QPI}(g) from Ref.~\citenum{wang_quasiparticle_2017}), we find that SOC induces a splitting of two QPI scattering vectors, which can be traced back to the avoided crossing of the $d_{xz/yz}$ and $d_{xy}$ bands observed near the Fermi level in Fig.~\ref{fig:ARPES}(e). A value of $\lambda=200$~meV reproduces the experimentally observed splitting of the dispersion best; see Fig.~\ref{fig:QPI}(g). This value of $\lambda$ also results in a simulated tunneling spectrum that is consistent with experiment: Fig.~\ref{fig:QPI}(h) shows the calculated SOC-induced gap (solid lines) for the same values of $\lambda$ as in panels (c-g), together with the experimental spectrum~\cite{marques_magneticfield_2021}.

Overall, this analysis reveals that very good agreement can be achieved from realistic modeling of QPI incorporating the DMFT self-energy. Here, the SOC-induced gap at the Fermi energy allows for quantitative assessment of the role of SOC at the surface of \ce{Sr2RuO4} and how it is modified by correlation effects, showing similar enhancement as in the bulk. 

\vspace{5mm}
\large\noindent\textbf{Signatures of superdispersion in tunneling spectroscopy}
\vspace{2mm}
\normalsize

\begin{figure}
    \centering
    \includegraphics[width=\linewidth]{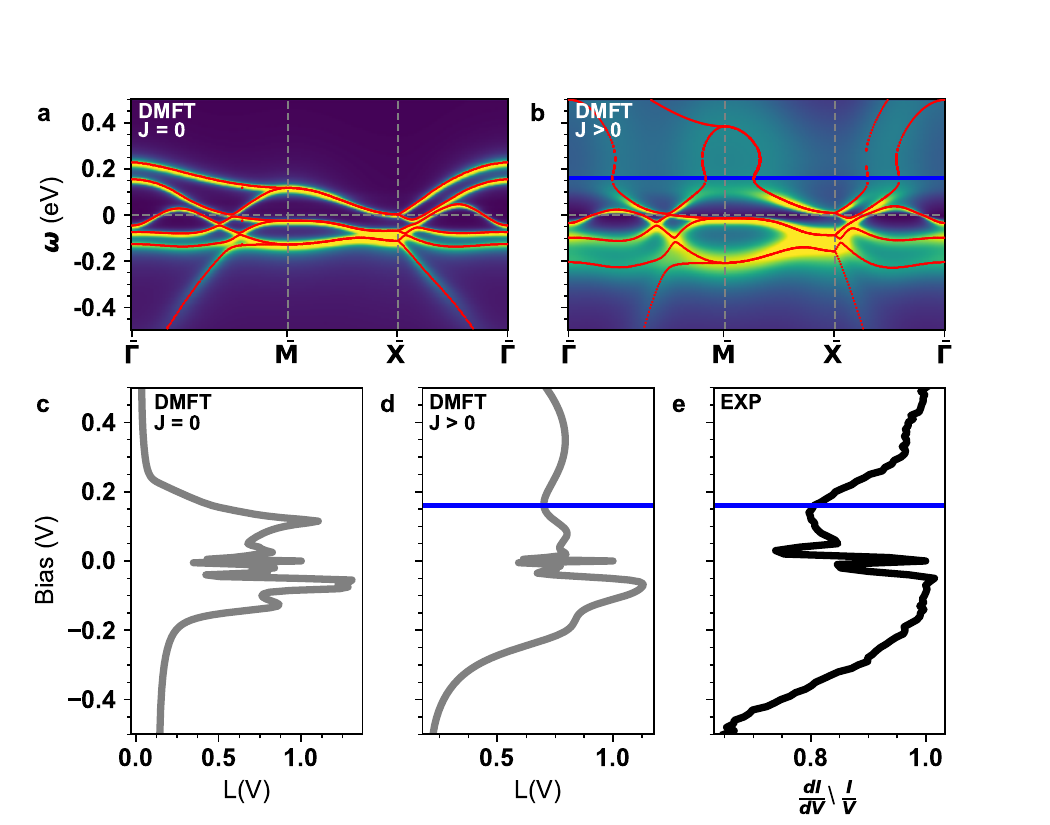}
    \caption{
        \textbf{Signatures of Hund superdispersion in tunneling spectroscopy.}
        (a, b)~Spectral function of the surface of \ce{Sr2RuO4} when incorporating the DMFT bulk self-energy obtained with (a) 
        artificial parameters $(U,J)=(4.5,0)~\mathrm{eV}$ and (b) 
        the physical parameters 
        $(U,J)=(2.3,0.4)~\mathrm{eV}$. 
        The parameters in (a) are chosen to produce similar quasiparticles weights as in (b).
        The difference between the two spectral functions demonstrates the influence of Hund coupling $J$. The red overlaid lines are the solutions to Eq.~\eqref{Eq:QPdisp} underlining the superdispersive feature at 160~meV in (b) with a change in curvature of the dispersion. (c, d)~Simulated tunneling spectra $L(V)$ [Eq.~\ref{feenstra}] for (c)~$J=0$ and (d)~$J>0$.
        (e)~Experimental tunneling spectrum $L(V)$ measured by STM ($V_s=500\mathrm{mV}$, $I_s=1\mathrm{nA}$, $V_L=8\mathrm{mV}$, $T=75-\mathrm{mK}$, see Methods). In both experiment and DMFT with $J>0$, a pronounced kink in the spectrum is observed around 160~meV (solid blue line in (b, d, e)), at the energy of the onset of the superdispersion. 
        } 
    \label{fig:DOS}
\end{figure}

Having established a framework that incorporates correlation effects and quantitatively reproduces both ARPES and STM experiments, we now turn our attention to the unoccupied states where the superdispersion is predicted to occur. To illustrate that the feature due to the superdispersion is a direct consequence of the Hund coupling in this system, in Figs.~\ref{fig:DOS}(a) and \ref{fig:DOS}(b), we plot the spectral function with the surface DFT band structure and DMFT bulk self-energy for zero and finite Hund coupling $J$. For $J=0$, we find a sharp quasiparticle dispersion with reduced band velocities. For $J>0$, we observe that the dispersion broadens significantly above $\sim 160~\mathrm{meV}$ (blue line in Fig.~\ref{fig:DOS}(b)), concomitant with a pronounced increase in $\hat{\Sigma}''$. By overlaying the solution to the pole equation [Eq.~\eqref{Eq:QPdisp}] in red, the superdispersive behaviour can be identified in Fig.~\ref{fig:DOS}(b) by the curvature inflection at $+160$~meV (blue line), which is absent in Fig.~\ref{fig:DOS}(a).  

To identify this spectral signature in tunneling spectroscopy, we calculate the continuum local density of states in a $\pm0.5$~eV energy window. To undertake this comparison with experirment for such a wide energy range, the energy dependence of the tunneling matrix elements cannot, in general, be neglected. As was shown by Feenstra~\cite{Feenstra_1994}, a useful way to normalize the spectrum such that the tunneling matrix elements cancel out is by considering
\begin{equation}
    L(\mathbf{r},V)=\frac{\rho(\mathbf{r},eV)}{\frac{1}{eV}\int_0^{eV}\rho(\mathbf{r},\omega)\mathrm{d}\omega} \approx  \frac{\frac{dI}{dV}(\mathbf{r},V)}{\frac{I(\mathbf{r},V)}{V}},
 \label{feenstra}
\end{equation}
where $\rho(\mathbf{r},eV)$ is the density of states at position $\mathbf{r}$ and energy $eV$ ($e$ is the electron charge), and $I$ and $V$ are the current and voltage, respectively. This quantity can be directly obtained in experiment from a measurement of the differential conductance $\frac{dI}{dV}$ and tunneling current $I$, and can be numerically calculated from $\rho(\mathbf{r},eV)$. We here consider $L(V)\equiv \langle L(\mathbf{r},V)\rangle_\mathbf{r}$ for a clean, defect-free surface. For $J=0$, the calculated Feenstra function obtained from a cLDOS calculation (Fig.~\ref{fig:DOS}(c)) is quickly suppressed above the quasiparticle band maximum at 200~meV. For $J>0$, the Feenstra function remains finite over a much larger energy range (Fig.~\ref{fig:DOS}(d)). Besides the sharp structure at the Fermi energy, it displays a pronounced dip at 160~meV, induced by the superdispersion and change in sign of $\partial_\omega {\Sigma}'_{mm}(\omega)$ (blue line in Fig.~\ref{fig:DOS}(d)). The agreement with the experimental $L(V)$ spectrum in Fig.~\ref{fig:DOS}(e) is remarkable: Not only is the spin-orbit gap at low-energies reproduced well, as seen in Figs~\ref{fig:QPI}(h) and \ref{fig:DOS}(e), but also the structure in a wider energy range. This includes a maximum at about $-150\mathrm{mV}$, with a decrease in $L(V)$ towards lower energies, but most importantly, the minimum at the same energy where the Hund superdispersion occurs. This feature remains robust against details of the band structure (e.g., an artificially imposed shift in the chemical potential), and it cannot be reproduced using a monotonous self-energy as obtained for $J=0$. It is thus a unique feature in the correlated density of states that does not have its origin in the bare band structure, but rather in the self-energy. It is a direct signature of the strong correlations induced by Hund coupling that drive the superdispersion.

\vspace{5mm}
\large\noindent\textbf{Discussion}
\vspace{2mm}
\normalsize

Our results demonstrate an integrated framework that combines DFT, DMFT, and continuum Green's function methods, enabling \textit{ab initio} calculations of tunneling spectra, QPI and ARPES for strongly correlated materials. So far, comparison of the spectral function between theory and experiment for such materials has been limited to ARPES. Previous QPI simulations used a fixed, phenomenological band renormalization~\cite{chandrasekaran_engineering_2024}, which works well in the immediate vicinity of the Fermi energy~\cite{kreisel_quasi-particle_2021}, but fails to capture, e.g., finite quasiparticle lifetimes or an energy-dependent band renormalization. Our improved method enables direct reproduction of both low- and high-energy structure of QPI from a single model.

In the specific case of \ce{Sr2RuO4}, we have demonstrated excellent agreement between theory and experiment with almost no free parameters. An intriguing result is the robustness of the bulk self-energy that we embedded into the surface. The comparison between theory and experiment suggests that using the bulk self-energy to describe correlations in the surface layer, where the atomic structure changes due to octahedral rotations, works remarkably well.
Such a good agreement is not trivially expected: one could have thought that the Lifshitz transition of the $d_{xy}$ vHs that occurs due to surface reconstruction would modify the dominant interactions. The $d_{xy}$ vHs moves from above the Fermi energy to below it, but the energy relative to the Fermi energy stays about the same, which rationalizes why the self-energy does not change more. Recent measurements of the bulk electronic structure by ARPES under uniaxial strain suggest that, while there is a notable difference of the correlation effects at the Lifshitz transition, the lifetime of the quasiparticles for the vHs just above or below the Fermi level is very similar~\cite{hunter_NFL2025}, supporting the conclusions presented here. 
One might expect that, in more isotropic systems, the self-energy at the surface becomes significantly modified, and hence treatments of a layer-dependence of the self-energy and the impact of the $k_z$ dispersion will become important~\cite{Rhodes_2019_kz,rhodes_nature_2023}.

Through comparison of the tunneling spectrum with state-of-the-art numerical simulations, our work provides direct experimental evidence of the Hund superdispersion, a long-standing theoretical prediction, validating the DMFT-derived self-energy in \ce{Sr2RuO4}.

A natural system to extend such studies is \ce{Sr2MoO4}. Indeed, in contrast to \ce{Sr2RuO4} where this signature of Hund metallicity occurs at positive energy 
because the $t_{2g}$ shell occupied by 4 electrons is more than half-filled, in \ce{Sr2MoO4} with 2 electrons the superdispersion is predicted in the occupied states~\cite{PhysRevLett.125.166401}. 
This would allow direct quantitative comparisons between photoemission, STM, and theory. Beyond transition metal oxides, Hund coupling plays an important role in iron-based superconductors~\cite{yin_kinetic_2011}, where similar signatures are expected, which would provide an important benchmark in establishing their correlated electronic structure and for the understanding of their superconductivity.

By enabling the inclusion of correlation effects in QPI simulations via the DMFT self-energy, our work unlocks the ability to accurately capture experimental results not only in the immediate vicinity of the Fermi energy, but also across the transition to incoherent electronic states on larger energy scales.
Therefore, it gives tunneling spectroscopy and QPI access to probe the electronic structure also in the unoccupied states, with exquisite resolution better than $1\mathrm{meV}$, as well as in magnetic fields. 
It enables one, e.g., to investigate how the electronic structure of a correlated material changes across a quantum critical point, where correlation effects are expected to dominate, and through theory--experiment comparisons distinguish different scenarios for quantum criticality~\cite{sachdev_topological_2019,yang_critical_2023,cui_deconfined_2025}, a hotly debated open question for many materials.


\vspace{5mm}
\large\noindent\textbf{Methods}
\vspace{2mm}
\normalsize

\textbf{DFT calculations} \\
We use tight-binding models both for the bulk and the surface of the material, obtained by projecting the results of DFT simulations (\textsc{Wien2k}~\cite{10.1063/1.5143061} for the bulk and Quantum Espresso~\cite{Giannozzi_2017} for the surface). For the bulk, we start from the experimental primitive cell in the $I4/mmm$ space group and relax the structural parameters. The bulk structure is used for the calculation of the self-energy via DMFT.
For the surface, we consider a free-standing layer of \ce{Sr2RuO4}, including the octahedral rotations as observed experimentally for the surface layer. For the octahedral rotation angle, we use the value determined experimentally from IV LEED~\cite{matzdorf_surface_2002}, which is consistent with that obtained from fitting the band structure~\cite{chandrasekaran_engineering_2024}.
In both bulk and surface cases, we project the Ru-$t_{2g}$ orbitals using \textsc{Wannier90}~\cite{Pizzi_2020}, add a local SOC and adjust the chemical potential to maintain charge neutrality.

\textbf{DMFT calculations} \\
Describing STM measurements requires calculations at low temperatures $<10~\mathrm{K}$ with fine low-energy resolution in real frequencies. 
These requirements are met by DFT+DMFT+NRG \cite{kugler_strongly_2020,Kugler2024,Grundner2025,Kugler2026,LaBollita2026}, i.e., 
using the numerical renormalization group (NRG)~\cite{Bulla2008} as the DMFT impurity solver with a tight-binding model downfolded from DFT. 
Including SOC or octahedral rotations in a three-orbital model with crystal-field splitting, however, exceeds the present capabilities of NRG.
Hence, we compute the bulk self-energy without SOC, while both SOC and octahedral rotations \textit{are} included in the band structure in the simulations.
Our DFT+DMFT+NRG setting is thus analogous to Ref.~\citenum{kugler_strongly_2020}, but with improved low-energy resolution thanks to a symmetric improved estimator for the self-energy~\cite{Kugler2022} and adaptive Brillouin-zone integration~\cite{Kaye2023,Van-Munoz/Beck/Kaye:2024}. 
The code uses the MuNRG package~\cite{Lee2021,Lee2017,Lee2016} built on top of the QSpace tensor library~\cite{Weichselbaum2012a, Weichselbaum2020, Weichselbaum2024}.

\textbf{cLDOS calculations} \\
To simulate STM measurements, we use the continuum Green's function technique~\cite{choubey_visualization_2014,kreisel_interpretation_2015,kreisel_quasi-particle_2021} as implemented in calcQPI~\cite{wahl_calcqpi_2025} to calculate the continuum local density of states (cLDOS). We insert $\hat{\Sigma}(\omega)$ in the orbital basis into the Green's function of the clean host,
\begin{equation}
    \hat{G}_{\Sigma}(\mathbf{k},\omega)=\left(\omega\cdot\mathbb{1}-\hat{H}_0(\mathbf{k})-\hat{\Sigma}(\omega)\right)^{-1},
    \label{Eq:GF-kspace}
\end{equation}
where $\hat{H}_0(\textbf{k})$ is the non-interacting tight-binding Hamiltonian in the orbital basis.
From Eq.~\eqref{Eq:GF-kspace}, we obtain the full lattice Green's function $\hat{G}(\mathbf{R},\mathbf{R^\prime},\omega)$  of the system including a point defect using the $T$-matrix formalism~\cite{economou_greens_2006}, 
\begin{equation}    
\hat{G}(\mathbf{R},\mathbf{R^\prime},\omega)=\hat{G}_{\Sigma}(\mathbf{R}-\mathbf{R}^\prime,\omega)+\hat{G}_{\Sigma}(\mathbf{R},\omega)\hat{T}(\omega)\hat{G}_{\Sigma}(-\mathbf{R}^\prime,\omega),
\end{equation}
where the $\hat{T}$-matrix is given by
\begin{equation}
    \hat{T}=\hat{V}\left(\mathbb{1}-\hat{V} \hat{G}_{\Sigma}(0,\omega)\right)^{-1}.
\end{equation}
From the lattice Green's function, we compute the continuum Green's function $G(\mathbf{r},\mathbf{r},\omega)$ and  the local density of states $\rho(\mathbf{r},\omega)=-\mathrm{Im}G(\mathbf{r},\mathbf{r},\omega)$ using the Wannier functions obtained from \textsc{Wannier90} and calcQPI~\cite{wahl_calcqpi_2025}. Here, $R$ describes a discrete lattice vector, and $r$ describes a continuous real-space vector. To first order, the differential conductance measured in STM is proportional to the local density of states, $dI/dV(\mathbf{r},V)\sim \rho(\mathbf{r},eV)$.

\textbf{Scanning tunneling microscopy} \\
Tunneling spectroscopy measurements in Fig.~\ref{fig:DOS}(e) were performed using a home-built STM mounted in a dilution refrigerator~\cite{singh_construction_2013} with a base temperature below $100\mathrm{mK}$. These measurements were performed with open feedback loop conditions after stabilizing the tip-sample distance at setpoint conditions ($I_\mathrm s$, $V_\mathrm s$) as specified in the figure captions. Spectra are acquired using a lock-in amplifier to measure the differential conductance ($dI/dV$) as a function of voltage $V$ with a voltage modulation $V_\mathrm{L}$ added to the bias.

\vspace{5mm}
\large\noindent\textbf{Acknowledgments}
\vspace{2mm}
\normalsize

We thank Roser Valent\'i, Vidya Madhavan, and Jernej Mravlje  for insightful discussions.
LCR and PW acknowledge funding from the Leverhulme Trust through Research Project Grant RPG-2022-315, as well as LCR through UKRI3345 and CM and PW through UKRI1107. PK acknowledges funding from EP/T02108X/1. This work has used computational resources of Archer2 in Edinburgh and the high-performance computing cluster Hypatia at the University of St Andrews.
FBK acknowledges funding from the Ministerium f\"ur Kultur und Wissenschaft des Landes Nordrhein-Westfalen (NRW-R\"uckkehrprogramm).
The Flatiron Institute is a division of the Simons Foundation. 

\vspace{5mm}
\large\noindent\textbf{Author contribution statement}
\vspace{2mm}
\normalsize

LCR led the simulations and analysis using the DMFT self-energies provided by FBK. CAM provided the STM data, and EM and PK the ARPES data. LCR performed the DFT calculations for the surface model. PW developed the code to include self-energies in QPI calculations. LCR, FBK, OG, AG, and PW wrote the manuscript. All authors discussed and contributed to the manuscript and analysis. PW initiated the project.

\vspace{5mm}
\large\noindent\textbf{Conflict of interest}
\vspace{2mm}
\normalsize

The authors declare no conflicts of interest.

\newpage

\renewcommand{\thefigure}{S\arabic{figure}}
\renewcommand{\theequation}{S\arabic{equation}}
\renewcommand{\thesection}{S\arabic{section}}

\setcounter{section}{0} 
\setcounter{figure}{0}
\setcounter{table}{0}
\setcounter{equation}{0}









\begin{center}
\large{Supplementary Material for \\ `Revealing Hund superdispersion with tunneling spectroscopy'}
\end{center}

\author{Luke C. Rhodes\,\orcidlink{0000-0003-2468-4059}}
\affiliation{SUPA, School of Physics and Astronomy, University of St Andrews, North Haugh, St Andrews, KY16 9SS, United Kingdom}

\author{Fabian B.~Kugler\,\orcidlink{0000-0002-3108-6607}}
\affiliation{Institute for Theoretical Physics, University of Cologne, 50937 Cologne, Germany}

\author{Olivier~Gingras\,\orcidlink{0000-0003-3970-6273}}
\affiliation{Center for Computational Quantum Physics, Flatiron Institute, 162 5th Avenue, New York, New York 10010, USA}
\affiliation{Université Paris-Saclay, CNRS, CEA, Institut de physique théorique, 91191, Gif-sur-Yvette, France}

\author{Carolina Marques\,\orcidlink{0000-0002-3804-096X}}
\affiliation{SUPA, School of Physics and Astronomy, University of St Andrews, North Haugh, St Andrews, KY16 9SS, United Kingdom}

\author{Edgar Abarca Morales\,\orcidlink{0000-0002-7714-8228}}
\affiliation{Max Planck Institute for Chemical Physics of Solids, Nöthnitzer Strasse 40, 01187 Dresden, Germany}

\author{Phil D.C. King\,\orcidlink{0000-0002-6523-9034}}
\affiliation{SUPA, School of Physics and Astronomy, University of St Andrews, North Haugh, St Andrews, KY16 9SS, United Kingdom}

\author{Antoine Georges\,\orcidlink{0000-0001-9479-9682}}
\affiliation{Collège de France, Université PSL, 11 Place Marcelin Berthelot, 75005, Paris, France}
\affiliation{Center for Computational Quantum Physics, Flatiron Institute, 162 5th Avenue, New York, New York 10010, USA}
\affiliation{Centre de Physique Théorique, Ecole Polytechnique, CNRS, Institut Polytechnique de Paris, 91128, Palaiseau Cedex, France}
\affiliation{DQMP, Université de Genève, 24 Quai Ernest Ansermet, 1211, Genève, Suisse}

\author{Peter Wahl\,\orcidlink{0000-0002-8635-1519}}
\affiliation{SUPA, School of Physics and Astronomy, University of St Andrews, North Haugh, St Andrews, KY16 9SS, United Kingdom}
\affiliation{Physikalisches Institut, Universität Bonn, Nussallee 12, 53115 Bonn, Germany}

\date{\today}


\section{Difference between the bulk and surface structures and high-symmetry points}

\begin{figure}[h!]
    \centering
    \includegraphics[width=0.7\linewidth]{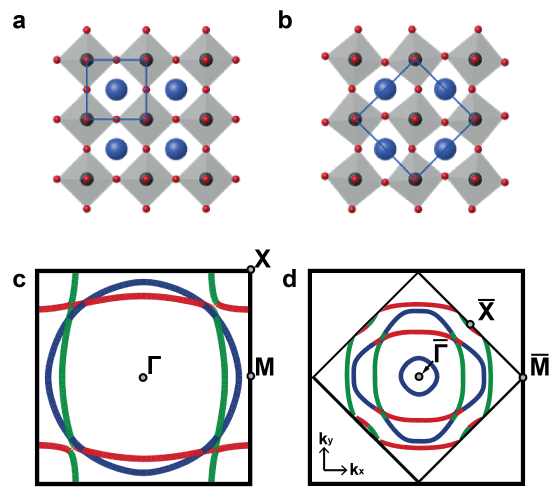}
    \caption{
        \textbf{Fermi surface and high-symmetry points of the tight binding models for the bulk and surface structures.} (a)~Atomic structure of the bulk. The blue square denotes the one-Ru atom unit cell. The Ru, O and Sr atoms are depicted in grey, red and blue, respectively.
        (b)~Similar figure for the surface structure, which has a two-Ru atom unit cell.
        (c)~Fermi surface on the Brillouin zone of the bulk structure, with high-symmetry points. The green, red and blue colors correspond to the characters associated to the $d_{xz}$, $d_{yz}$ and $d_{xy}$, respectively.
        (d)~Similar figure for the surface structure. The Fermi surfaces in (c,d) are plotted on the same momentum scale. Note the change in position of the high-symmetry points.
        The bulk structure (a,c) is used in Fig.~1 of the main text. 
        The surface structure (b,d) is used in Figs~2, 3, 4 of the main text. 
    } 
    \label{fig:supp-BrillouinZones}
\end{figure}

\ce{Sr2RuO4} has a well documented surface relaxation \cite{matzdorf_surface_2002,marques_magneticfield_2021} that freezes out a in-plane rotational phonon mode resulting in a doubling of the unit cell, as depicted in Fig.~\ref{fig:supp-BrillouinZones}(a,b). The details regarding how this modifies the electronic structure are discussed in the main text. For clarity, in Fig.~\ref{fig:supp-BrillouinZones}(c,d), we plot the 2D Fermi surface and high-symmetry points for each corresponding unit cell.

\begin{figure}
    \centering
    \includegraphics[width=\linewidth]{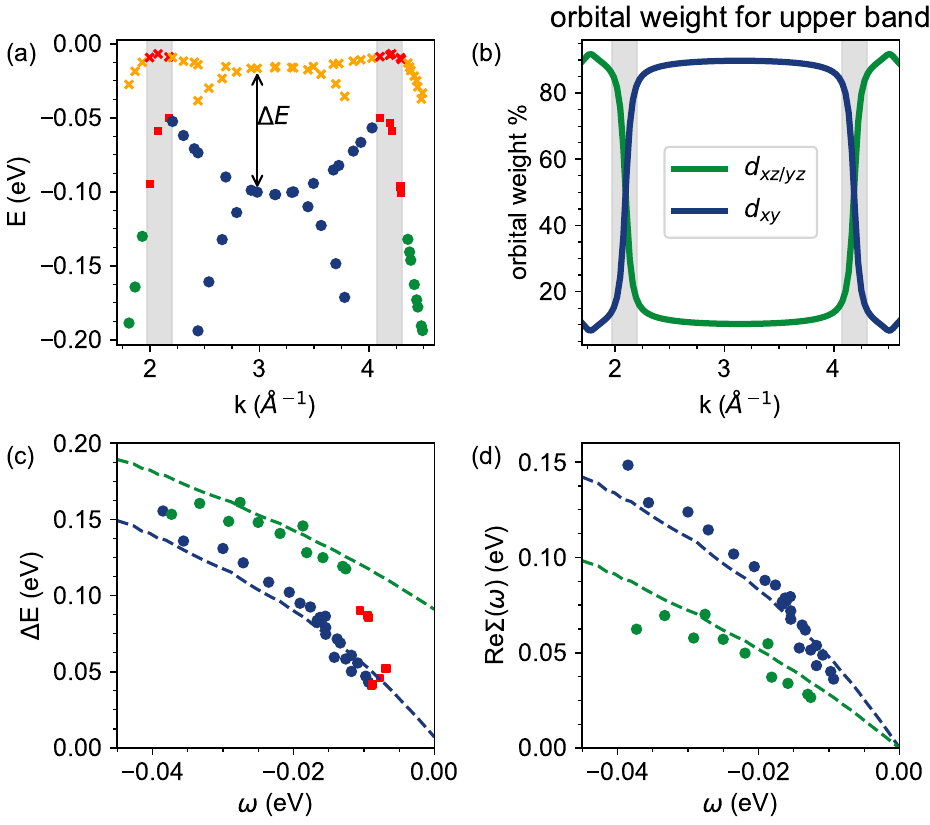}
    \caption{
        \textbf{Extracting the self-energy from ARPES}. 
        (a)~ARPES data points extracted from Ref.~\citenum{Morales_Hierarchy_2023,chandrasekaran_engineering_2024} around the $\bar{\text{M}}$ point of \ce{Sr2RuO4}, where the surface band corresponding to a van Hove singularity can be observed. The orange crosses are experimental data points, the blue and green circles are the eigenvalues obtained from the DFT calculation using the surface electronic structure of \ce{Sr2RuO4}, as discussed in Fig.~\ref{fig:supp-BrillouinZones}. The red points in (a) correspond to points of mixed orbital character and have thus been neglected.
        (b)~Orbital mixing along the same \textbf{k}-point path. Spin-orbit coupling creates an hybridization gap between the $d_{xy}$ (blue) and $d_{xz/yz}$ (green) bands, which mixes the orbital weight significantly close to band crossings.
        (c)~The difference in energy between the calculated and experimental states for fixed momenta, plotted against the experimental energy. The blue and green dotted lines corresponds to the self-energy of the bulk of \ce{Sr2RuO4} minus the chemical potential shift required for particle conservation.
        (d)~The same as (c), neglecting the red points, and rigidly shifting the green and blue data such that the real part of the self-energy becomes zero at $\omega=0$, for comparison with Ref.~\citenum{tamai_high-resolution_2019}.
    }
    \label{fig:supp-ARPES_self_energy}
\end{figure}

\section{Extraction of self-energy from ARPES}
To estimate the self-energy from angle-resolved photoemission spectroscopy (ARPES) data, we employ an equivalent method to that performed by Tamai \textit{et al.}~\cite{tamai_high-resolution_2019}. We begin by assuming that the non-interacting DFT band structure and experimental measurement are related via 
\begin{equation}
    E_\mathrm{exp}(\mathbf{k}) = E_\mathrm{DFT}(\mathbf{k}) -  \mathrm{Re}\Sigma(\mathbf{k},\omega),
\end{equation}
such that the difference, $\Delta E$, is related to the real part of the self-energy, $\mathrm{Re}\Sigma(\mathbf{k},\omega)$, as shown in Fig.~\ref{fig:supp-ARPES_self_energy}(a). Because $E_\mathrm{exp}(\mathbf{k})$ for the surface bands is only available in a limited energy range and a limited part of the Brillouin zone, we have to restrict our analysis to parts of the band structure that are dominated by a single orbital character. The analysis is therefore carried out on a specific cut around the van Hove singularity (vHs) at the M point which has two regions where this is true: the $d_{xz/yz}$ dominated outer bands (shown in green) and the $d_{xy}$ dominated vHs (shown in blue). These bands strongly mix orbital characters at certain momenta, as shown by the grey shaded regions in Fig.~\ref{fig:supp-ARPES_self_energy}(b) and the red points in Fig.~\ref{fig:supp-ARPES_self_energy}(a). These points are discarded in the analysis of the self-energy. In Fig.~\ref{fig:supp-ARPES_self_energy}(c), we plot $\Delta E(k)$ against $E_\mathrm{exp}(\mathbf{k})$ to obtain the experimental part of the self-energy, separated by orbital character. The dotted lines are the DMFT calculated self-energy for the given orbital character with the chemical potential subtracted. The points shown in red have been discarded. Finally, in order to more closely compare our conclusions with the analysis of Ref.~\citenum{tamai_high-resolution_2019}, we rigidly subtract an offset $\mathrm{Re}\Sigma(\mathbf{k},\omega=0)$ from the points for the $d_{xy}$ and $d_{xz/yz}$ bands in (c). The result is shown in Fig.~2(d) of the main text.

\newpage

\label{Bibliography}

\bibliographystyle{unsrtnat}
\bibliography{references}

\end{document}